# Local Interpretability of Calibrated Prediction Models: A Case of Type 2 Diabetes Mellitus Screening Test

Simon Kocbek[†]
Faculty of Health Sciences
University of Maribor
Maribor Slovenia
Advanced Analytics Institute
University of Sydney
Sydney Australia
skocbek@gmail.com

Primoz Kocbek, Leona Cilar
Faculty of Health Sciences
University of Maribor
Maribor Slovenia
primoz.kocbek@um.si,
leona.cilar1@um.si

Gregor Stiglic
Faculty of Health Sciences,
Faculty of Electrical Engineering
and Computer Science
University of Maribor
Maribor Slovenia
gregor.stiglic@um.si

## ABSTRACT

Machine Learning (ML) models are often complex and difficult to interpret due to their "black-box" characteristics. Interpretability of a ML model is usually defined as the degree to which a human can understand the cause of decisions reached by a ML model. Interpretability is of extremely high importance in many fields of healthcare due to high levels of risk related to decisions based on ML models. Calibration of the ML model outputs is another issue often overlooked in the application of ML models in practice. This paper represents an early work in examination of prediction model calibration impact on the interpretability of the results. We present a use case of a patient in diabetes screening prediction scenario and visualize results using three different techniques to demonstrate the differences between calibrated and uncalibrated regularized regression model.

## 1 Introduction

Interpretability is the degree to which a human can understand the cause of a machine learning (ML) decision [18]. Generally higher interpretability correlates with better understanding and trust in model predictions [19]. Usually we classify interpretability of ML models as Model-Specific or Model-Agnostic. Model-specific interpretation is limited to specific types of models (e.g., the interpretation of regression weights in a linear model) and is usually intrinsic to the model. Model-agnostic interpretation is more general and can be applied to any ML model [7], but is usually done post-hoc. Interpretability in ML models is often classified in global and local interpretability. Global interpretability refers to understanding how the model works globally by inspecting parameters of a complex model. On the other hand, local interpretability refers to an individual prediction of a model specific for a single instance from the dataset [6].

Many predictive ML models in healthcare provide effective predictions but in practice, healthcare experts still need good reasoning to be able to trust in the ML based decisions [15]. To ensure that reliable and trustworthy ML models are used, those need to be validated.

Various calibration techniques have been developed in order to improve the probability estimation in terms of the error distribution of an existing ML model. The evaluation of those models is crucial prior to their application in practice [1]. The process of calibrating a prediction model is defined as "learning a function that maps the original probability estimates, or scores, into more accurate probability estimates". Common metrics used to estimate the goodnes of fit after calibration are the Binomial test, the Brier Score Loss, the Chi-squared test, the Traffic Lights Approach, and the Hosmer-Lemeshow test [8].

In this paper, we present a single patient use case to demonstrate possible impact the calibration of ML models might have on their interpretability at the local level. We use three different types of local interpretability visualizations to explain the predictions of a LASSO regression model in 10-years Type 2 Diabetes Mellitus (T2DM) prediction. Results are reported only for one ML technique, since the experiments showed that no calibration was needed for other two techniques.

## 2 Data and Methods

### 2.1 Data

The Survey of Health, Ageing and Retirement in Europe (SHARE) data [3] was used to conduct the experiments in this study. SHARE represents a multidisciplinary and cross-national database of data on health of about 140,000 individuals aged 50 or older. The dataset is freely available and contains data from different European countries. Participants of the SHARE survey from Austria, Belgium, Denmark, France, Germany, Greece, The Netherlands, Israel, Italy, Spain and Sweden, Switzerland were included in this study. Although it focuses on many aspects of ageing the SHARE data also contains a wide range of explanatory features that may have impact on the onset of self-reported T2DM.

Data from SHARE survey waves 1 to 7 collected between 2004 and 2017 were used in this study to develop and validate prognostic models for 10-year T2DM prediction in European countries [3]. Israel was the only exception, where wave 2 data were collected in 2009-2010 period and therefore the prediction window was shortened by approximately two years. The data are freely available to registered researchers from the SHARE Research Data Center (http://www.share-project.org).

The features included in our experiments consisted of four groups as follows: 2 Demographic features (Age and Gender), 15 Physical health features (e.g., Long-term illness, High blood pressure or hypertension, etc.), 31 Aggregated health data features (e.g., Body Mass Index (BMI), Quality of life, etc.), and 5 Behavioural data features (e.g., Smoking, Alcohol consumption, etc.).

The target feature in our experiments was the presence of T2DM diagnosis in the follow-up period of 10 years for all participants who reported no T2DM diagnosis at the baseline interview at wave 1 or wave 2. The following question was used to define the T2DM status: "Diabetes or high blood sugar: ever diagnosed?". In case of participants who were included in the study in wave 1, we checked for presence of self-reported T2DM in all consecutive waves up to wave 6. Wave 7 was used as an endpoint for participants who entered the study in wave 2. The mean difference between the first (wave 1 or 2) and the last (wave 6 or 7) interview was $10.5 \pm 0.7$ years. Out of 16,363 participants, eligible for the model development phase, 1,360 (8.3 %) answered positively to the diabetes/high blood sugar question in waves 6/7.

In Table 1, we summarize features of interest for our experiments.

**Table 1**: Summary table for features of interest in 10-year T2DM prediction

|  | T2DM* (8.3%, n=1,360) | No T2DM* (91.7%, n= 15,003) |
|---|---|---|
| Age | 65.4 (SD=8.8) | 64.9 (SD=9.2) |
| Female | 52.7% (n=717) | 44.5% (n=6674) |
| BMI | 28.8 (SD=4.6) | 26.1 (SD=4) |
| EverSmokedDaily | 50.5% (n=687) | 47.6% (n=7142) |
| AlcoholConsumption | 4.6 (SD=2.3) | 4.2 (SD=2.2) |
| Sport | 2.6 (SD=1.3) | 2.4 (SD=1.3) |
| HighBloodPressure | 42.6% (n=579) | 30.8% (n=4615) |
| HighCholesterol | 26.7% (n=363) | 20.8% (n=3125) |
| Orientation | 3.8 (SD=0.5) | 3.9 (SD=0.4) |
| MathSkills | 3.4 (SD=1.1) | 3.5 (SD=1.1) |
| QualityOfLife | 36.6 (SD=6.1) | 37.9 (SD=5.8) |
| SelfPercievedHealth2 | 76.5% (n=1040) | 65.4% (n=9817) |
| Guilt | 6.9% (n=94) | 7.4% (n=1115) |

*followup period of 10 years

The selected "use-case patient" for whom we performed local interpretation was a T2DM positive 59 years old male, with:
- BMI of 28.04,
- history of high blood pressure (HighBloodPressure = 1),
- no information about high cholesterol (HighCholesterol = 0),
- no alcohol consumption in the last six months (AlcoholConsumption = 7),
- reported long term illness (LongTermIlness = 1),
- hardly ever active with sports (Sport = 4),
- high math skills (MathSkills = 5),
- measured orientation of 3 out of 0 - 4 (Orientation =3),
- quality of life of 47 (QualityOfLife = 47), for more information please see [13], and
- self perceived health of less than very good health (SelfPerceivedhealth2 = 1).

## 2.2 Methods

Multiple binary classification models were used with a focus on relatively simple models used in practice, both linear, such as Least Absolute Shrinkage and Selection Operator (LASSO) [9] and nonlinear, such as Random Forest, XGBoost [5]. The data was split 80:20 for training and testing, and 10 times repeated 10-fold cross validation was used for model performance evaluation (AUC). For local interpretation, we performed manual split of the data, where 100 instances were used for testing the interpretability, and 16,262 for training the model.

After initial model development, the model output was calibrated using a linear regression model and visually compared to uncalibrated model. We used flexible loess calibration curves with pointwise 95% confidence intervals and provided summary statistics for the calibration slope for the overall effect of the predictors and the calibration intercept as in [4]. We would like to note, that in case of non-linear models the calibration was not needed based on the visual inspection, and we report results only for LASSO.

For interpretation of the model we used the DALEX (Descriptive mAchine Learning EXplanations) package [2], which offers a wide range of methods for local interpretability ( i.e., with respect to observations). In our experiments we used iBreakDown [10] visualization where contributions of features are calculated sequentially with the effects depending on the change of expected model prediction while all previous features are fixed. Additionally we used SHAP (SHapley Additive exPlanations) [16] values based on the coalitional game theory, where the contribution of the feature is calculated as an average of contributions of each possible ordering of features. We also applied individual profiles (Ceteris-paribus Profiles) visualization showing prediction model outcomes for a use-case patient for different values of a numeric feature to gain better insight into model behaviour [19]. All experiments were implemented in R language and environment for statistical computing [20].

## 3 Results

Figure 1 illustrates probability plots for the uncalibrated and calibrated models. It can be observed that the calibrated model fits the expected predicted vs. observed values much better than the uncalibrated model. The predictive performance of both models was the same with the AUC of 0.702 (95% CI: 0.699 - 0.706).

We manually inspected local interpretability plots to observe the difference in iBreakDown, SHAP and individual profile visualizations of calibrated vs. uncalibrated models. In general, predicted probabilities from the calibrated model were closer to mean predicted response.

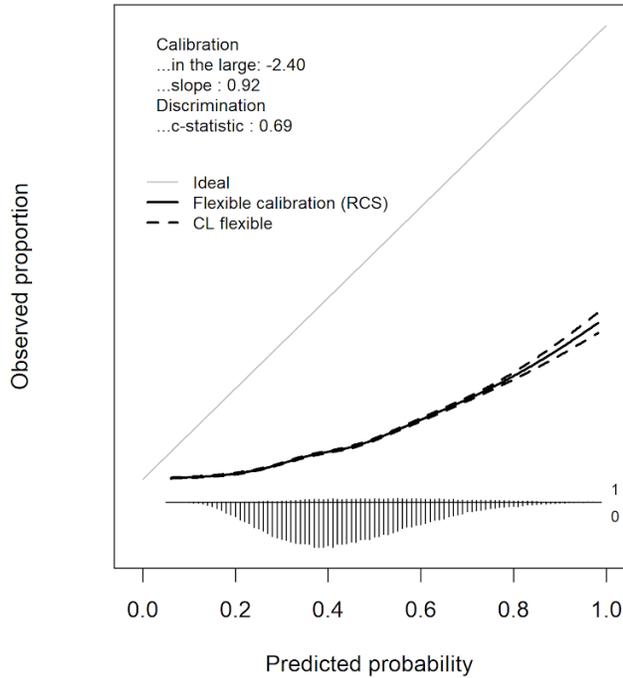
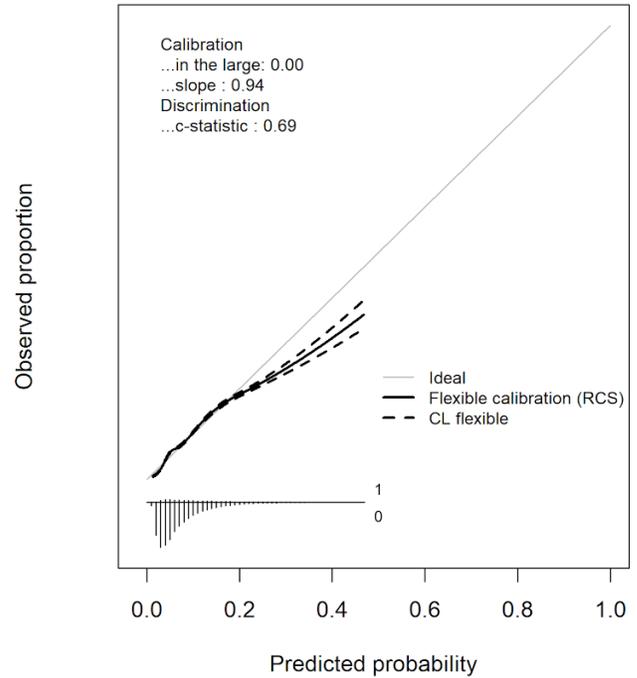

Figure 1: Comparison of probability plots for uncalibrated (left) and calibrated (right) model.

Since we identified some instances where some important features were differently ranked in calibrated vs. uncalibrated models, we illustrate one such example in Figure 2. Positive (green) and negative (red) local contributions for each feature are visualized using DALEX iBreakDown visualization. We notice that top 5 ranked features do not differ, while some differences can be observed in the features ranked in positions 6 to 10.

Perhaps most interestingly BMI is ranked at position 6, but not in the top 10 for the calibrated model, which could significantly influence the interpretation of the results for the healthcare experts. In Figure 3, we illustrate SHAP values for both the uncalibrated features and the calibrated model. SHAP is more resistant to order of features than iBreakDown, however, BMI is again missing in the uncalibrated model.

calibrated vs. the uncalibrated model. For individuals similar to the use case in this paper, BMI values lower than 30 would result in lower risk with the calibrated vs. the uncalibrated model. This changes for BMI values between 30 and 40 where the uncalibrated model predicts the higher risk level. It needs to be noted that this could also happen due to higher variance due to lower number of patients at the extreme values of the BMI.

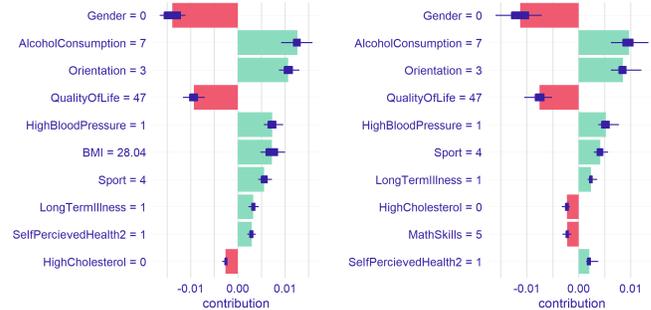

Figure 3: SHAP values for both models, uncalibrated left and calibrated right.

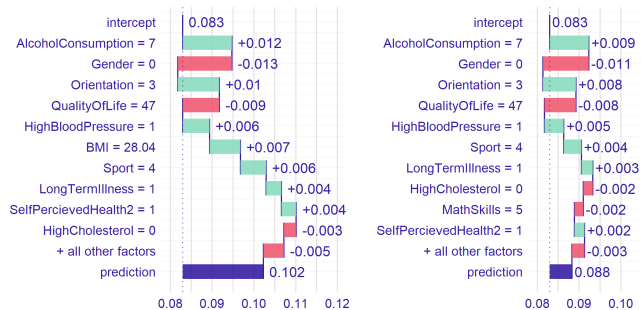

Figure 2: Localized feature importance for a selected positive instance, using the iBreakDown method, classified with uncalibrated (left) and calibrated (right) model.

We also observed individual profiles (Figure 4) for the outcome in relation to BMI, where minor differences can be seen for the

## 4 Discussion and Conclusion

In this paper, we present a use-case to demonstrate possible impact of calibration on interpretability of prediction models with focus on predicting 10-years risk of T2DM in older population. We built models with Random Forest, XGBoost and LASSO. Since Random Forest and XGBoost predictions resulted in well calibrated outputs, we further examined interpretability of calibrated and uncalibrated LASSO models.

Local interpretation showed that features that are commonly identified as T2DM risk factors such as alcohol consumption [14] or BMI [11] were ranked high in both iBreakdown and SHAP interpretation at least for uncalibrated model. Interestingly, age was not amongst top ranked features. This presents a difference with usual studies on T2DM prediction [12,17], where the population is not only elderly as in our study - hence, age does not have a big influence, if any.

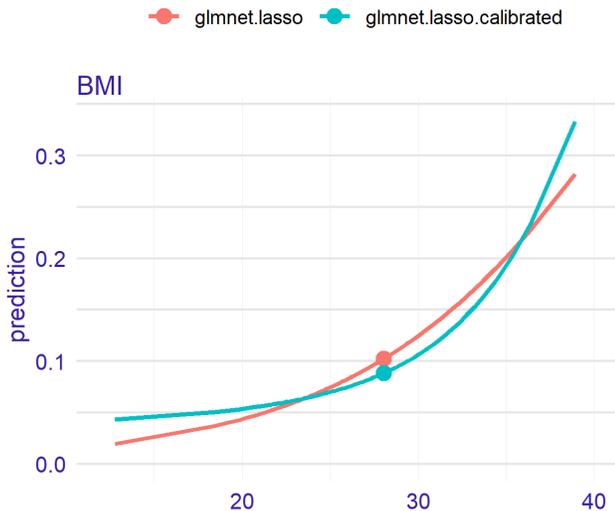

**Figure 4: Individual profiles for the uncalibrated and calibrated model.**

Results also showed that although differences in local interpretability between calibrated and uncalibrated models are rather small, it is possible that changes are evident in local interpretation visualizations and might influence the interpretation of the results by the healthcare experts.

Specifically, for our selected instance the BMI was not featured among top 10 most influential features while it was ranked at position 6 for the uncalibrated model. The individual profile plot for BMI also shows higher predicted risk provided by uncalibrated model for a patient with a BMI of 28.04. Such BMI value would be considered as a serious risk for the onset of T2DM and therefore we prefer the output of the uncalibrated model in this case. This is important since different conclusions can be made when ML models are interpreted by medical experts.

For future work, we would like to investigate more instances and their local interpretation with selected methods. This would be possible once an automated framework for inspection of differences between calibrated and uncalibrated models is developed. Therefore, this might also represent one of the challenges for the future.

## ACKNOWLEDGMENTS

This work was supported by the Slovenian Research Agency grants ARRS N2-0101 and ARRS P2-0057.